# Dynamical Constraints on Multi-Planet Exoplanetary Systems

Jonathan Horner [1], Robert A. Wittenmyer [1], Chris G. Tinney [1], Paul Robertson [2], Tobias C. Hinse [3, 4] and Jonathan P. Marshall [5]

[1] *Department of Astrophysics and Optics, School of Physics, University of New South Wales, Sydney, NSW, 2052, Australia*
[2] *Department of Astronomy and McDonald Observatory, University of Texas at Austin, Austin, TX 78712, USA*
[3] *Korea Astronomy and Space Science Institute, 776 Daedeokdae-ro Yuseong-gu 305-348 Daejeon, Korea*
[4] *Armagh Observatory, College Hill, Armagh, BT61 9DG*
[5] *Departmento Física Teórica, Facultad de Ciencias, Universidad Autónoma de Madrid, Cantoblanco, 28049 Madrid, Spain*

**Summary:** As a direct result of ongoing efforts to detect more exoplanetary systems, an ever-increasing number of multiple-planet systems are being announced. But how many of these systems are truly what they seem? In many cases, such systems are announced solely on the basis of orbital fits to observational data, and no attempt is made to see whether the proposed orbits are actually dynamically feasible. As a result, it is certain that planetary systems are being announced that involve planets moving on orbits that would be dynamically unstable on timescales of just a few hundred years.

Here, we present the results of dynamical simulations that investigate the orbital stability and evolution of a number of recently discovered exoplanetary systems. These simulations have enabled us to create highly detailed dynamical maps of those systems, allowing us to better constrain the orbits of the planets contained therein. In some cases, our results have even led to the very existence of the planets themselves being called into question.

**Keywords:** Planetary systems, Numerical methods: N body simulation, Planetary systems: dynamical evolution and stability, Exoplanets, Circumbinary companions

## Introduction

The question of whether we are alone in the universe is one that has been pondered since humanity first looked up at the night sky. Until the latter part of the twentieth century, such speculation remained firmly entrenched in science fiction, as the tools necessary to search for planets around other stars were beyond us. In the late 1980s and early 1990s, that changed, and astronomers began to search for planets around other stars. The first planetary system detected beyond our own Solar system was that around the pulsar PSR B1257 +12 (discovered in 1992) [1][2], a discovery which was quickly followed (in 1995) by the detection of the first planet orbiting a sun-like star, 51 Pegasi b [3]. Since those first discoveries, a vast number of exoplanets have been detected[1], at an ever-accelerating rate.

---

[1] As of 2[nd] November, 2012, the Extrasolar Planets Encyclopaedia, http://exoplanet.eu/, lists a total of 843 planets, whilst the stricter Exoplanet Data Explorer (http://exoplanets.org) lists a total of 632 planets as having been confirmed to date.

Planets have been found moving on extreme orbits, and around a diverse sample of stars (e.g. [4][5][6][[7][8][9]), and the announcement of new planets has essentially become routine.

When the first planets were found around other stars, the discovery process followed the maxim "extraordinary claims require extraordinary evidence" and, as such, every other possible explanation for the observed "planetary" signal was examined in detail, prior to the researchers feeling confident in announcing the planetary discovery. In recent years, as the presence of planets around a wide variety of stars has been confirmed, the rigour with which such claims are examined prior to announcement has, in some cases, diminished.

The detection of stars with multiple planets provides the opportunity for us to apply tools initially developed to study the behaviour of objects in our own Solar system to test whether proposed planetary systems are truly feasible. It is worth noting that the process of detecting multiple planet systems is often far more involved than the discovery of single planets. Such discoveries typically require observations to be carried out over multiple-year baselines before the presence of additional planets can be detected in a given system (e.g. [13][14]). As such, studying the dynamics of multiple-planet systems is an even younger science than the study of exoplanets themselves, which might explain why dynamical studies are often overlooked in papers announcing new multiple planet systems.

In this work, we present the results of dynamical studies of a wide variety of proposed multiple-planet systems in order to show how such dynamical tests can be used to form a critical component of the exoplanetary discovery process. In some cases, our results reveal that the planetary system, as proposed in the discovery work, is simply not dynamically feasible, and that there must be some other explanation for the observed signal. In others, our results can be used to obtain better constraints on the true nature of a detected planetary system.

In section 2, we describe the methodology used in our dynamical studies, before providing examples of how dynamical studies can force a significant rethink of the nature of proposed planetary systems in section 3. In section 4, we show how dynamical studies of other systems have helped to constrain their orbital architecture, cases where observational and theoretical work go hand in hand. Finally, in section 5, we present our conclusions.

## Dynamically Testing Exoplanetary Systems

In order to test the dynamical stability of multiple-planetary systems, we use the Hybrid integrator within the *n*-body dynamics package MERCURY [10]. It allows long-term dynamical integrations of a given exoplanetary system to be carried relatively quickly, whilst accurately integrating the effects of close encounters between any objects in the integration. This latter point is vital, if one is to determine the true dynamical stability of a given planetary system.

For each system we study, we follow a now well-established routine. We place the planet with the best-determined orbit on the best-fit orbit detailed in the relevant literature. We then create a suite of dynamical simulations in which the second planet in the system starts on a wide variety of initial orbits, distributed in even steps across the full $\pm 3\sigma$ error ellipse for a number of the planet's orbital elements. The precise number of integrations we carry out for a given exoplanetary system has changed since our first work, as we have gained access to more powerful supercomputing systems. In our earliest work, studying the proposed planets around HR8799 [11] and the cataclysmic variable HU Aquarii [12], we were only able to study of the order of ten thousand potential architectures at any given time. More recently, we have be able to expand our work so that we now consider approximately ten times as many trial systems, allowing a greater resolution across the orbital parameter space we consider.

We follow the dynamical evolution of each model system for a period of 100 Myr[2]. If the system becomes unstable during that time period, it will typically result in the catastrophic disintegration of the system via collisions, or the ejection of one or more bodies. We record the time at which such events happen, having set the ejection distance such that if the orbit of either planet is dramatically perturbed, it is considered ejected from the system. Typically, we consider the ejection of a planet to have occurred once it reaches a barycentric distance of 10 AU, since most of the objects studied initially move on orbits at just a couple of AU from the central star. In a couple of cases, where the orbit of one or more of the proposed planets was particularly extreme to begin with, this ejection criterion was relaxed to a slightly greater distance, in order that it only be triggered if the planet's orbit had been significantly altered.

The direct result of our integrations is that, for each individual orbital architecture considered, we obtain a lifetime – the time until the system being simulated fell apart. If the system survived for the full 100 Myr of our simulations, this lifetime is set at 100 Myr. This then enables us to create plots of the mean lifetime of the system as a function of the initial orbital elements considered for the second planet. An example of such a plot can be seen in Fig. 1, which shows the dynamical stability of the HD 142 system [13] as a function of the initial orbital eccentricity and semi-major axis of the candidate planet HD 142 d. For ease of comparison between the systems studied, all plots in this paper will show mean lifetimes on a scale ranging from $10^3$ to $10^8$ years, as in Fig. 1. In that plot, and all others in this work, red areas denote regions of significant dynamical stability, while yellows and blues show regions with progressively greater levels of instability. In the case of HD 142 d, our results show that the candidate planet is certainly dynamically feasible. They also provide an interesting additional constraint on the orbit of that planet, revealing that sufficiently high orbital eccentricities, although allowed when fitting orbits to the observational data, clearly yield unstable solutions.

---

[2] Simulations that run for a timescale of order 100 Myr are a good compromise for tests of the orbital stability of a given planetary system, since they allow the runs to complete in a reasonable amount of time, whilst also allowing us to examine the orbital stability on timescales that are a significant fraction of the lifetime of the exoplanet host stars.

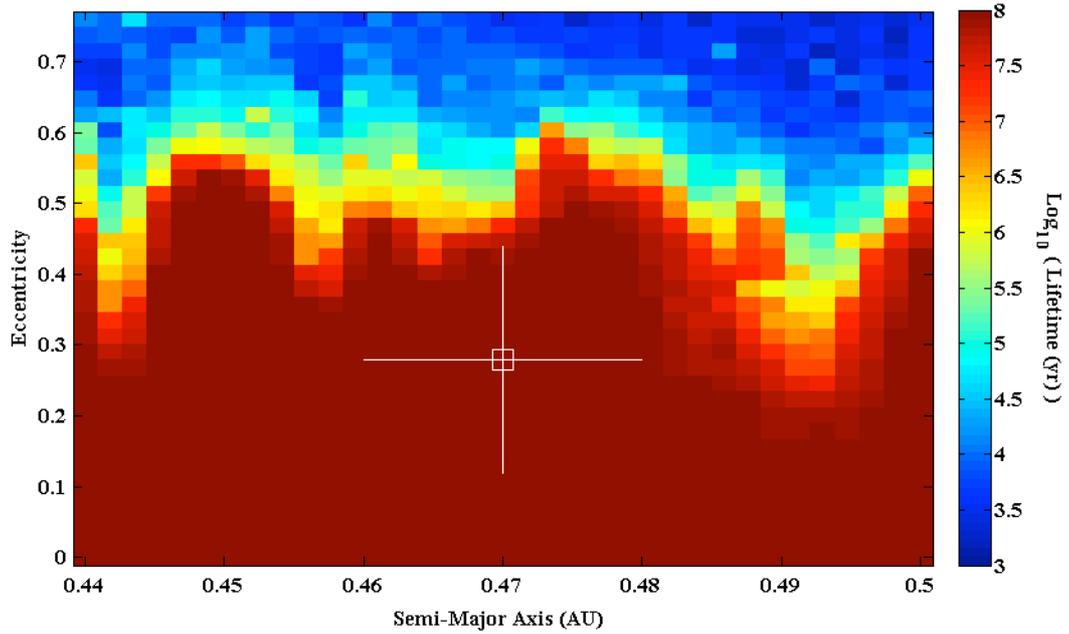

*Fig. 1: The dynamical stability of the HD142 planetary system, as a function of the initial semi-major axis and eccentricity of the orbit of HD142 d, a candidate planet of mass 0.3 times that of Jupiter. Each grid point shows the mean lifetime of 25 independent simulations carried out at that particular combination of initial semi-major axis and eccentricity, on a logarithmic scale. The box in the centre of the plot shows the nominal best-fit orbit for HD142 d, while the arms radiating from that point show the 1σ errors on that orbital solution. As can be seen, in this case, the nominal best-fit orbit for the planet lies in a region of extreme stability (mean lifetimes of 100 Myr, the maximum duration of our integrations). However, a broad region of extreme instability can also be seen, towards higher orbital eccentricities.*

## Dynamically Unfeasible Planetary Systems – the Circumbinary Planets

Perhaps the most important role played by dynamical simulations of newly announced exoplanetary systems is to test whether those systems are actually dynamically feasible. Although it might appear obvious that dynamical tests of any proposed system should form an integral part of the planet discovery process, many groups sadly neglect this aspect, choosing instead to publish proposed planetary systems on best-fit orbits that are often dynamically unrealistic. Discoveries of planets in one particular area have historically proven particularly prone to this mistake – the announcement of planets orbiting evolved binary star systems, such as HU Aquarii (e.g. [12][15][16][17]) and HW Virginis (e.g. [18][19]), discoveries based on variations in the timings of eclipses between the two binary components. In both discovery works, the proposed planets moved on best-fit orbits that were mutually crossing, and yet were not mutually resonant. The results of our dynamical simulations of these systems are shown in Fig. 2.

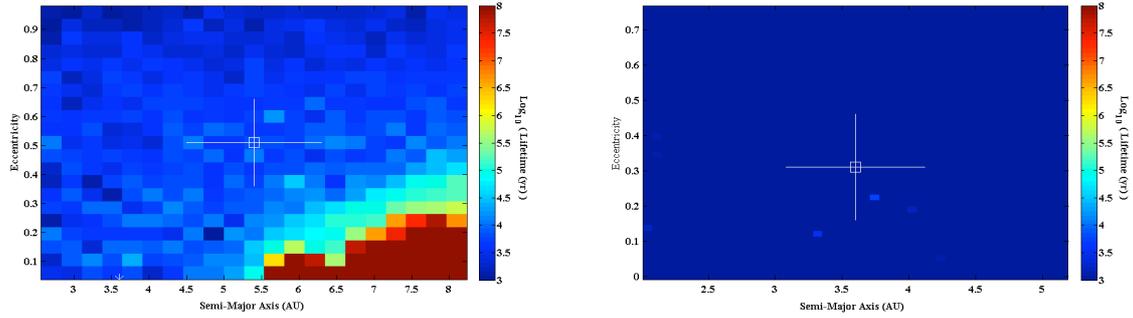

*Fig. 2: The dynamical stability of the proposed HU Aquarii (left) and HW Virginis (right) planetary systems. In both cases, the best-fit orbits lie in regions of extreme instability, with the systems being disrupted on timescales of hundreds or thousands of years.*

For both HU Aquarii and HW Virginis, we find that the planetary systems, as proposed in the discovery works ([15] and [18], respectively), are simply not dynamically feasible. In both cases, the proposed systems are dynamically unstable on timescales of hundreds or thousands of years – a result that calls into question the existence of planets in those systems, and most likely reveals that the observed eclipse timing variations are the result of a physical process other than planetary perturbations.

The recently proposed planetary system in the evolved eclipsing binary star system NSVS 14256825 ([20]) is another interesting case in point. In that work, the authors present observations of the eclipsing binary system taken over a period of several years, and showed that the timing of the eclipses between the two stars therein was deviating from the predicted times in a systematic fashion. Following the hypothesis that the variation was the result of perturbations from unseen giant planets in the system, they found that the variation could be explained by the presence of two giant planets, significantly more massive than Jupiter, moving on mutually-crossing orbits strikingly similar to those proposed to explain the similar timing variations observed for the HW Virginis and HU Aquarii systems. It therefore seemed prudent to carry out a detailed dynamical study of the system, following our earlier work on circumbinary planetary systems ([12], [16], [19], [21]). The results of our dynamical analysis of this system can be seen in Fig. 3.

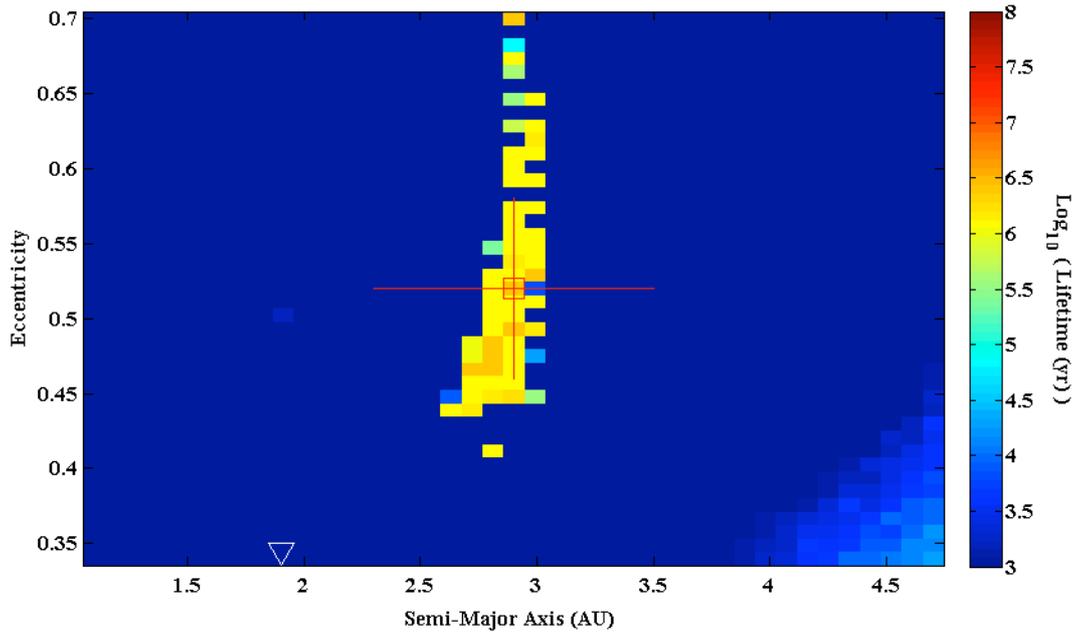

*Fig. 3: The dynamical stability of the proposed NSVS 14256825 planetary system, as a function of the orbital eccentricity and semi-major axis of proposed planet NSVS 14256825 c. For reference, the location of the nominal semi-major axis for the best-fit orbit for NSVS 14256825 b is denoted by the white triangle (the eccentricity of that proposed planet's orbit was fixed at 0.0 in the discovery work). Due to the difficulty seeing white lines on a yellow background, the 1σ errors for this system are shown in red.*

It is clear from Fig. 3 that, just like the planets proposed to orbit HU Aquarii and HW Virginis, the planets proposed to orbit the evolved binary system NSVS 14256825 do not stand up to dynamical scrutiny. The system is incredibly unstable, regardless of the initial orbit considered for NSVS 14256825 c, aside from a narrow region of moderate stability in the centre of the plot. Even in this region, however, mean lifetimes are only of the order of a few million years. Indeed, of the 126,075 independent trials carried out for this system, only 56 remained stable for the full 100 million years of our integrations – in fact, only 67 of the tested systems survived for more than one million years!

The fact that the proposed planetary systems around HW Virginis, HU Aquarii and NSVS 14256825, as discussed in the discovery papers, are so strikingly similar (featuring an inner planet on a circular orbit and an outer planet with an orbital period almost double that of the innermost, moving on a highly eccentric orbit) suggests that there may be some common property of these systems that is mimicking the presence of planets. This conclusion is reinforced when one notes that, in each case, the periods of the planetary orbits fall in the range ~3 to ~15 years, with the period of the outermost proposed planet being approximately double that of the inner[3]. Indeed, the same situation is seen for the proposed planets in the NN Serpentis system ([23]), although the proposed orbits for those planets do seem to lie in a region of dynamical stability ([21], see also Fig. 4). Given how similar all these proposed planetary systems are, and that, of the four systems studied, three are utterly dynamically unfeasible, it is almost certain that there is something other than planets causing the observed eclipse timing variations. Such close binary systems are particularly difficult to model – particularly those (such as HU Aquarii) in which a significant amount of matter is continually being transported from one star to the other. It is quite likely that some physical process within the binary itself is likely causing the observed timing variations, with two dominant frequencies in an approximate ratio 2:1. Whilst attempting to determine the nature of that process is beyond the scope of this work, it will certainly present an interesting avenue for future work.

---

[3] One further close binary system, SZ Herculis, also appears to display very similar behavior, which was in the past attributed to the presence of two low-mass M-dwarf companions, moving on highly eccentric orbits close to mutual 2:1 resonance [22]. Just as for the case of the proposed circumbinary planets discussed in this work, [31] have shown that the proposed low-mass companions to SZ Herculis are also dynamically unfeasible.

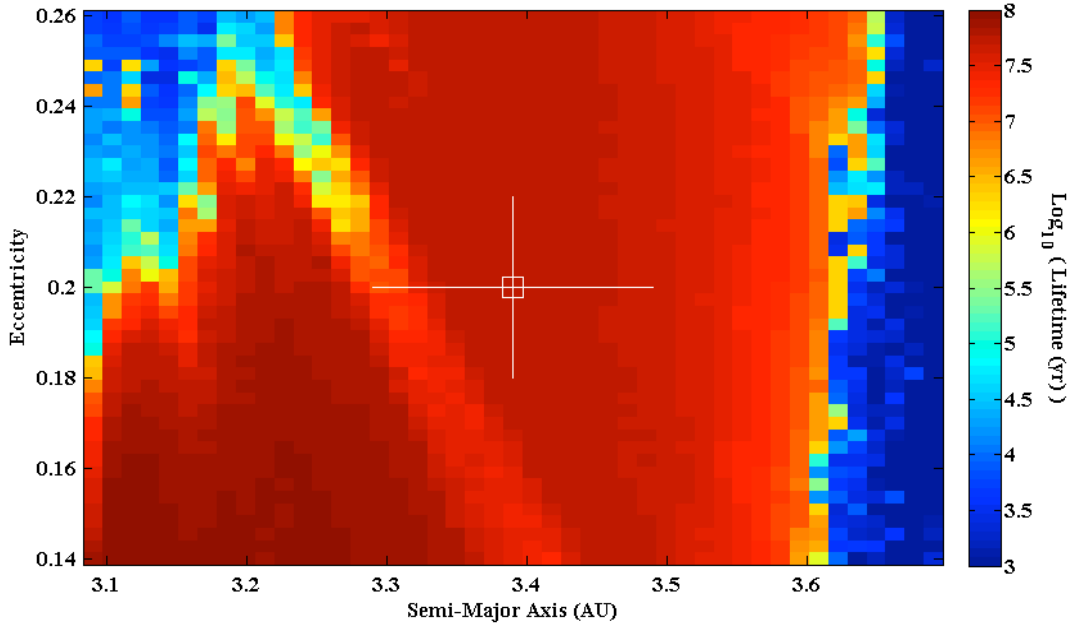

*Fig. 4: The dynamical stability of the proposed NN Serpentis planetary system, based on the orbits put forward in [23], for their 2:1 resonant solution. Unlike the other proposed circumbinary planets studied in this work (Figs. 2 and 3), the proposed planets are dynamically stable, so long as they move on orbits protected from close encounters by mutual 2:1 mean motion resonance.*

## Constraining the architecture of new multiple-planet systems

Dynamical studies of newly discovered multiple-planet systems do not always yield results that call the existence of the proposed planets into question. In many cases, such studies instead support the existence of the proposed planets, and can even help to better constrain their orbits. One such case is clearly seen in Fig. 1, where we show that the candidate planet HD 142 d moves on an orbit that is dynamically stable, so long as the orbital eccentricity is not too great. In that case, the nominal best-fit orbit lies within a broad region of strong dynamical stability, with the only unstable solutions lying beyond the 1σ errors on the planet's orbital solution.

The case for dynamical studies helping to constrain the orbits of proposed exoplanets is perhaps best exemplified for cases where those planets are proposed as moving on mutually resonant orbits – as is the case for the planets discovered in the HD155358 ([5]) and HD204313 ([24]) systems.

The discovery of the planetary system orbiting the star HD155358 system was first announced in 2007 ([25]). At that time, the authors noted that the two proposed planets were dynamically interacting, although the proposed orbits were stable for periods of at least $10^8$ years. The orbits proposed in that work had the planets relatively widely separated, with orbital periods of 195 and 530 days, respectively. The authors noted that there was some ambiguity regarding the period of the outer of the two planets, HD 155358c, with a second solution, featuring an orbital period of ~330 days, also allowing a reasonable fit to the observed data. As a result, the system was revisited in 2012 ([5]), at which point 51 further measurements of the radial velocity of HD 155358 were available for analysis. With this improved dataset, analysis yielded a significantly better two-planet solution for the HD 155358 system, with planets b and c moving on orbits with period 194.3 and 391.9 days – a period ratio of almost exactly 2:1. As is well known from studies of our own Solar system (e.g. [26], [27], [28]), objects moving on orbits with mutually commensurate periods can be either extremely dynamically stable, or extremely unstable. Stable scenarios are the result of mutual mean-motion resonance between the two objects of interest, where the commensurability prevents mutual close-encounters on astronomically long timescales. As such, it seemed prudent to study the dynamical stability of the proposed HD 155358 system, in order to ascertain whether the 2:1 commensurability proposed was dynamically stable, or unstable. Our results can be seen in Fig. 5.

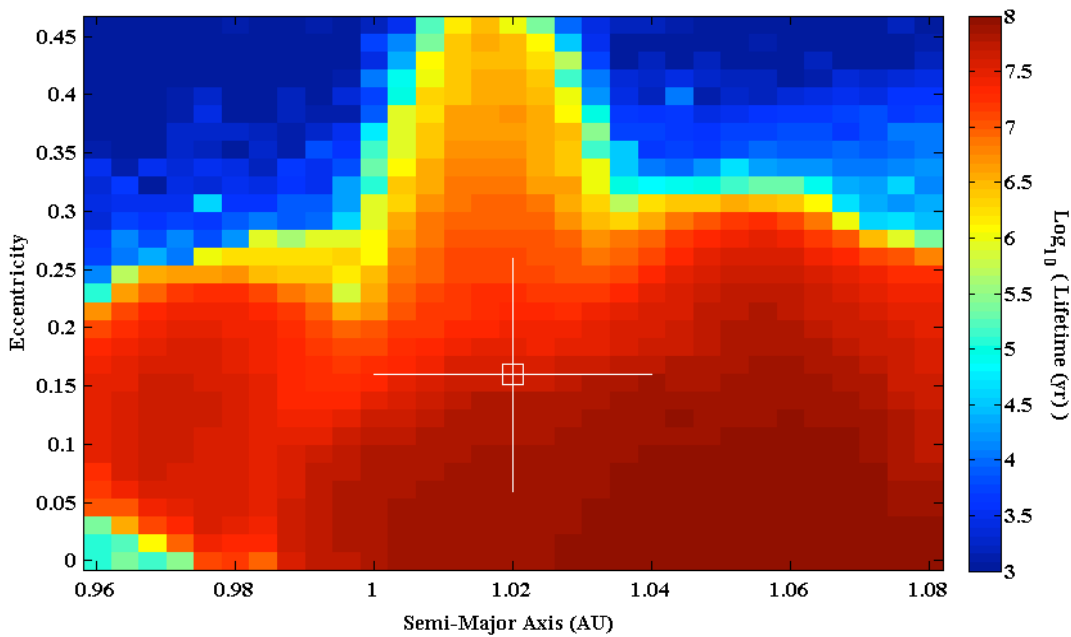

*Fig. 5. The dynamical stability of the HD 155358 planetary system, as a function of the semi-major axis and eccentricity of the orbit of HD 155358 c.*

It is immediately apparent that the nominal best-fit orbit for HD 155358 c lies in a broad region of dynamical stability, with very stable solutions stretching throughout the region within 1σ of the nominal best-fit orbit in both eccentricity and semi-major axis. As was the case for HD 142, we can see that low eccentricity solutions typically offer far greater prospects for stability than high eccentricity solutions, which helps to further constrain the orbit of HD 155358 c. The influence of the mutual 2:1 mean-motion resonance between planets HD 155358 b and c is clearly visible, offering a region of enhanced stability up to high eccentricities. Due to the strong protective nature of the resonant motion, even mutually crossing orbits can be dynamically stable (as seen in our own Solar system for a wide range of resonant objects, including the Jovian and Neptunian Trojans (e.g. [27], [29]) and the Plutinos (e.g. [28]).

As with HD 155358, the planetary system proposed around HD 204313 ([24]) features two planets moving on orbits that are close to, or trapped in, mutual mean-motion resonance. In this case, the planets lie strikingly close to mutual 3:2 resonance – the same resonance occupied by Pluto and the Plutinos in our own Solar system ([28]). Once again, given the potentially resonant nature of the HD 204313 system, it seemed prudent to study its dynamical stability, to see whether the proposed system was truly dynamically feasible. Our results can be seen in Fig. 6.

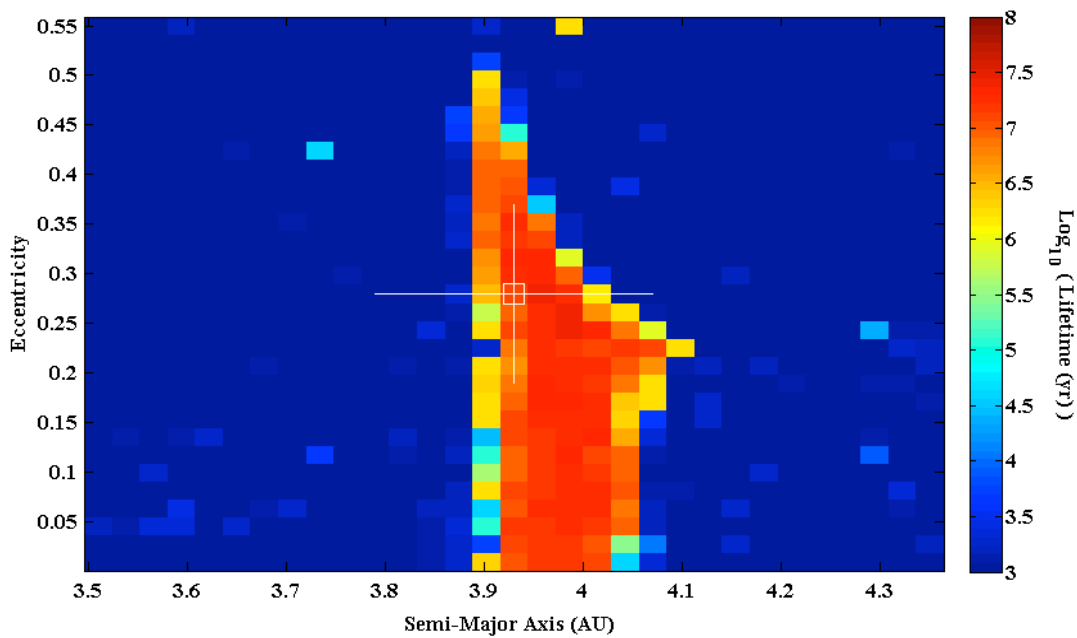

*Fig. 6. The dynamical stability of the HD 204313 planetary system, as a function of the semi-major axis and eccentricity of the orbit of HD 204313 c.*

As was the case for the planets in the HD 155358 system, we find that the nominal best-fit orbit for HD 204313 c lies right on top of a region of strong dynamical stability. Unlike that system, however, the region of stability is narrow, and is surrounded on all sides by a sea of strong instability. The region of stability offered by the 3:2 mean motion resonance is narrow, and sculpted at high eccentricities. Interestingly the distribution of stable orbits in Fig. 6 is strongly reminiscent of the Solar system's Plutino population. This is a prime example of how dynamical studies can offer significant improvements in the precision with which the orbits of exoplanets are known. The planets in the HD 204313 are only dynamically feasible if they are currently trapped in mutual 3:2 mean-motion resonance – otherwise the system falls apart on timescales of just a few thousand years.

A final example of a resonant exoplanetary system for which it is possible to better constrain the orbital architecture through dynamical integrations is HD 200964 ([30]). The planets proposed in that work have orbital periods of 613.8 days and 825.0 days, suggesting that they are trapped in mutual 4:3 mean-motion resonance. The authors carried out some small-scale dynamical integrations of the system, but noted that their simulation results could not yet conclusively confirm that the planets were in resonance, and urged further, more detailed dynamical investigation. We therefore chose to carry out such an investigation, the results of which can be seen in Fig. 7.

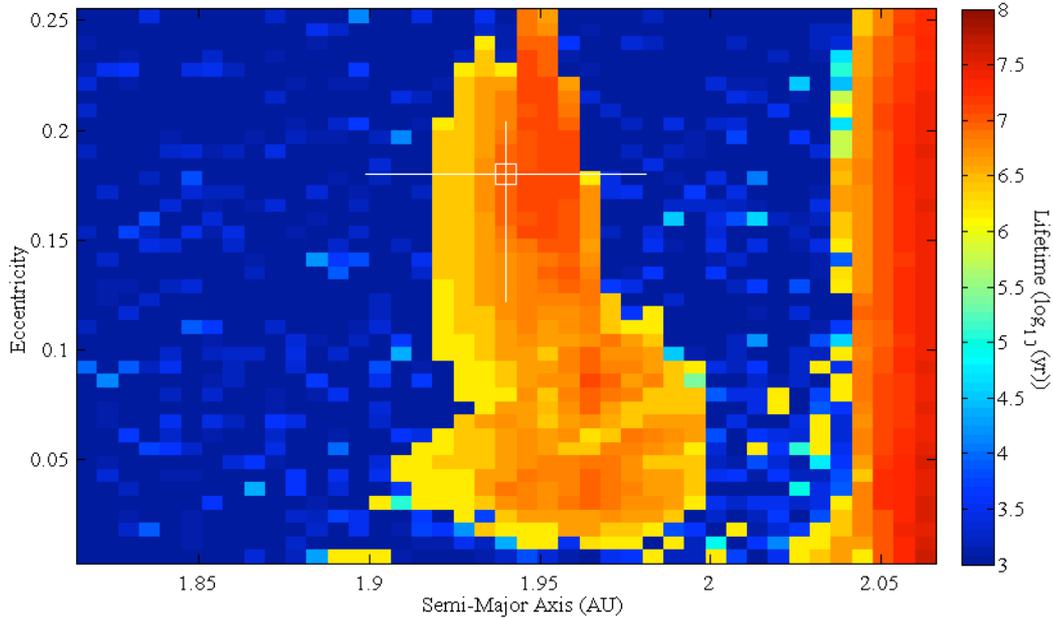

*Fig. 7. The dynamical stability of the HD 200964 planetary system, as a function of the semi-major axis and eccentricity of the orbit of HD 200964c.*

As was the case for the planets in the HD 204313 and HD 155358 systems, the nominal best-fit solution for HD 200964 c falls right in the middle of an island of stability, surrounded by a wide range of highly unstable orbital solutions. As such, it is almost certain that HD 200964 b and c are trapped in their mutual 4:3 mean-motion resonance. Interestingly, however, our simulations reveal a second broad region of stability, at semi-major axes of ~2.05 AU or greater. That result is not unexpected – and is a common result of such dynamical stability studies. Once HD 200964 c is placed beyond 2.05 AU, it is sufficiently far from HD 200964 b that stable non-resonant solutions are feasible. For smaller separations, the two planets strongly perturb one another, leading to extreme dynamical instability on very short timescales, unless the planets happen to be protected from such instability by mutual mean-motion resonance.

## Conclusions

The discovery of planets around other stars is a costly and time-consuming process. Observations of stars are taken over a period of many years, and are studied to search for any periodic variations that might be explained by the presence of planets orbiting those stars. In the early days of exoplanetary science, any suspected planetary detection was subject to rigorous investigation, prior to being announced. In recent years, however, as the detection of exoplanets have moved from the extraordinary to the mundane, a bewildering diversity of exoplanets are being announced orbiting an extremely diverse sample of stars. It is now the case that the presence of planets is regularly invoked to explain periodic variations observed in distant stars.

In this work, we show how detailed dynamical simulations can be used to examine proposed exoplanetary systems, and determine whether those systems are truly dynamically feasible. In many cases, we find that the proposed planets simply **do not work**, moving on orbits that are unstable on timescales of just hundreds or thousands of years – far shorter than the billion-year lifetimes of their host stars. In these cases, we suggest that another explanation is needed for the observed periodic variations in the host star or stars. If there are exoplanets orbiting those stars, they must move on drastically different orbits to those proposed in the discovery works – although it seems more likely that some other physical process is creating the observed variations.

We also show how dynamical studies can help astronomers to refine and better constrain the orbital architecture of proposed multiple-planetary systems. In the case of the planets discovered in the HD 155358, HD 204313 and HD 200964 systems, for example, our results reveal that the proposed planets are only dynamically stable if they move on mutually resonant orbits, and that a wide range of the observationally allowed orbital solutions are not dynamically feasible. In each of these cases, it is reassuring that the nominal best-fit orbital solutions for the planets in question lie in the middle of islands of stability, a case where detailed dynamical studies add further evidence for the presence of planets in those systems.

In the future, such dynamical studies will form a crucial part of the exoplanet discovery and announcement process. In principle any new multiple-planetary system should be studied dynamically to see whether the proposed planets truly make sense. We are now routinely carrying out such dynamical studies, as part of the Anglo-Australian Planet Search, where any suspected multiple-exoplanet systems detected are dynamically investigated prior to being announced.

## Acknowledgements

The work was supported by iVEC through the use of advanced computing resources located at the Murdoch University, in Western Australia. This research has made use of NASA's Astrophysics Data System (ADS), and the SIMBAD database, operated at CDS, Strasbourg, France. This research has also made use of the Exoplanet Orbit Database and the Exoplanet Data Explorer at exoplanets.org. The authors also wish to thank the two referees of this paper, Dr. Elliott Koch and Brett Addison, for their comments, which helped to improve the flow and clarity of our article.